\documentclass[prl,preprint,showpacs,superscriptaddress]{revtex4-1}
\usepackage[english]{babel}
\usepackage[latin1]{inputenc}
\usepackage{subfigure}
\usepackage{amsxtra} 
\usepackage{amsmath}
\usepackage{amssymb}
\usepackage{amstext}
\usepackage{amsthm}
\usepackage{amsfonts}
\usepackage{graphicx}
\usepackage[rigidchapters,outermarks]{titlesec}
\usepackage{fancyhdr}
\usepackage{courier}
\usepackage{natbib}
\usepackage{braket}
\usepackage{fourier}
\date{\today}

\begin{document}
\title{Macrospin Dynamics in Antiferromagnets Triggered by Sub-20 femtosecond Injection of Nanomagnons}
\author{D. Bossini$^{\star}$} \email[]{d.bossini@science.ru.nl}
\affiliation{Radboud University, Institute for Molecules and Materials, Heyendaalseweg 135, Nijmegen, The 
Netherlands}
\author{S. Dal Conte$^{\star}$}
\affiliation{Dipartimento di Fisica, Politecnico di Milano, Piazza Leonardo da Vinci 32, Milano, Italy}
\affiliation{Istituto di Fotonica e Nanotecnologie, Consiglio Nazionale delle Ricerche, Piazza Leonardo da Vinci 32, Milano, Italy}
\author{Y. Hashimoto}
\author{A. Secchi}
\affiliation{Radboud University, Institute for Molecules and Materials, Heyendaalseweg 135, Nijmegen, The 
Netherlands}
\author{R. V. Pisarev}\affiliation{A. F. Ioffe Physical-Technical Institute, Russian Academy of Sciences, 194021 St. Petersburg, 
Russia}
\author{Th. Rasing}\affiliation{Radboud University, Institute for Molecules and Materials, Heyendaalseweg 135, 
Nijmegen, The Netherlands}
\author{G. Cerullo}
\affiliation{Dipartimento di Fisica, Politecnico di Milano, Piazza Leonardo da Vinci 32, Milano, Italy}
\affiliation{Istituto di Fotonica e Nanotecnologie, Consiglio Nazionale delle Ricerche, Piazza Leonardo da Vinci 32, Milano, Italy} 
\author{A. V. Kimel}\affiliation{Radboud University, Institute for Molecules and Materials, Heyendaalseweg 135, 
Nijmegen, The Netherlands}

${\star}$ These authors equally contributed to the work

\maketitle

{\bfseries{The understanding of how the sub-nanoscale exchange interaction evolves in macroscale correlations and ordered phases of matter, such as magnetism and superconductivity, requires to bridge the quantum and classical worlds. This monumental challenge\cite{Polkovnikov2011,Aoki2014} has so far only been achieved for systems close to their thermodynamical equilibrium\cite{Antropov1995,Antropov1996}. Here we follow in real time the ultrafast dynamics of the macroscale magnetic order parameter triggered by the impulsive optical generation of spin excitations with the shortest possible nanometer-wavelength and femtosecond-period. Our experiments also disclose a possibility for the coherent control of these femtosecond nanomagnons, which are defined by the exchange energy. These findings open up novel opportunities for fundamental research on the role of short-wavelength spin excitations in magnetism and high-temperature superconductivity, since they provide a macroscopic probe of the femtosecond dynamics of sub-nanometer spin-spin correlations and, ultimately, of the exchange energy. With this approach it becomes possible to trace the dynamics of such short-range magnetic correlations for instance during phase transitions. Moreover, our work suggests that nanospintronics and nanomagnonics can employ phase-controllable spin waves with frequencies in the 20 THz domain.}}

Experimental studies allowing to investigate correlated matter in general, and magnetism in particular, at the length and time-scales of the exchange interaction have recently developed into an exciting research area. The experiments involving ultrashort timescales provided intriguing results, like the femtosecond laser induced transient ferromagnetic state of a ferrimagnet alloy\cite{Radu2011} and even superconductivity\cite{Fausti2011}. However, in these cases the wavelengths of the photo-induced excitations lie orders of magnitude above the nanometer length scale of the exchange interaction. An alternative strategy consists in the investigation of magnetic order induced by introducing impurities with atomic resolution in space\cite{Loth2012,Khajetoorians2012,Khajetoorians2013}, but these are static experiments at equilibrium. A fundamentally new approach to the problem, that combines the femtosecond timescale and nanometer lengthscale consists in studying the ultrafast dynamics of macroscale magnetic order parameter triggered by spin excitations with wavelength and period pertinent to the length- and time-scales of the exchange interaction.
 
These spin excitations correspond to magnons (or spin waves) with wavevector near the edges of the Brillouin zone. In antiferromagnetic materials such magnons can be elegantly excited in the time domain, via a second order impulsive stimulated Raman scattering (ISRS) process involving pairs of magnons with wavevectors almost equal in magnitude and opposite in sign. Although this process is allowed throughout the whole Brillouin zone, the magnon density of states is largest in the high-frequency region near the zone edges, which is dominated by the exchange interaction\cite{Cottam1986,Fleury1968} (see Supplementary Fig. S1). Thus a bound state of two high-energy, high-wavevector and counter-propagating spin waves, usually denoted as \textit{two-magnon} (2M) \textit{mode}, can be induced by a femtosecond light pulse. The frequency and wavevector of this magnetic excitation are the sums of the frequencies and wavevectors of the two magnons involved in the bound state\cite{Fleury1968,Cottam1986,Balucani1973,Cottam1972}. Although an impulsive excitation of such 2M mode was reported, the subsequent dynamics of the magnetic order parameter has not even been discussed yet\cite{Zhao2004}.  
 
In this Letter, we disclose the fastest possible dynamics of the macroscopic order parameter in a magnetic system, by means of an impulsive all-optical injection and detection of short-range spin excitations near the edges of the Brillouin zone. The wavelengths of the corresponding magnons are of the order of 1 nm and the frequencies in the 20 THz range. We demonstrate that the phase of the spin waves can be controlled by changing the polarization of the excitation beam.

\begin{figure}[h]
	\centering
	\includegraphics{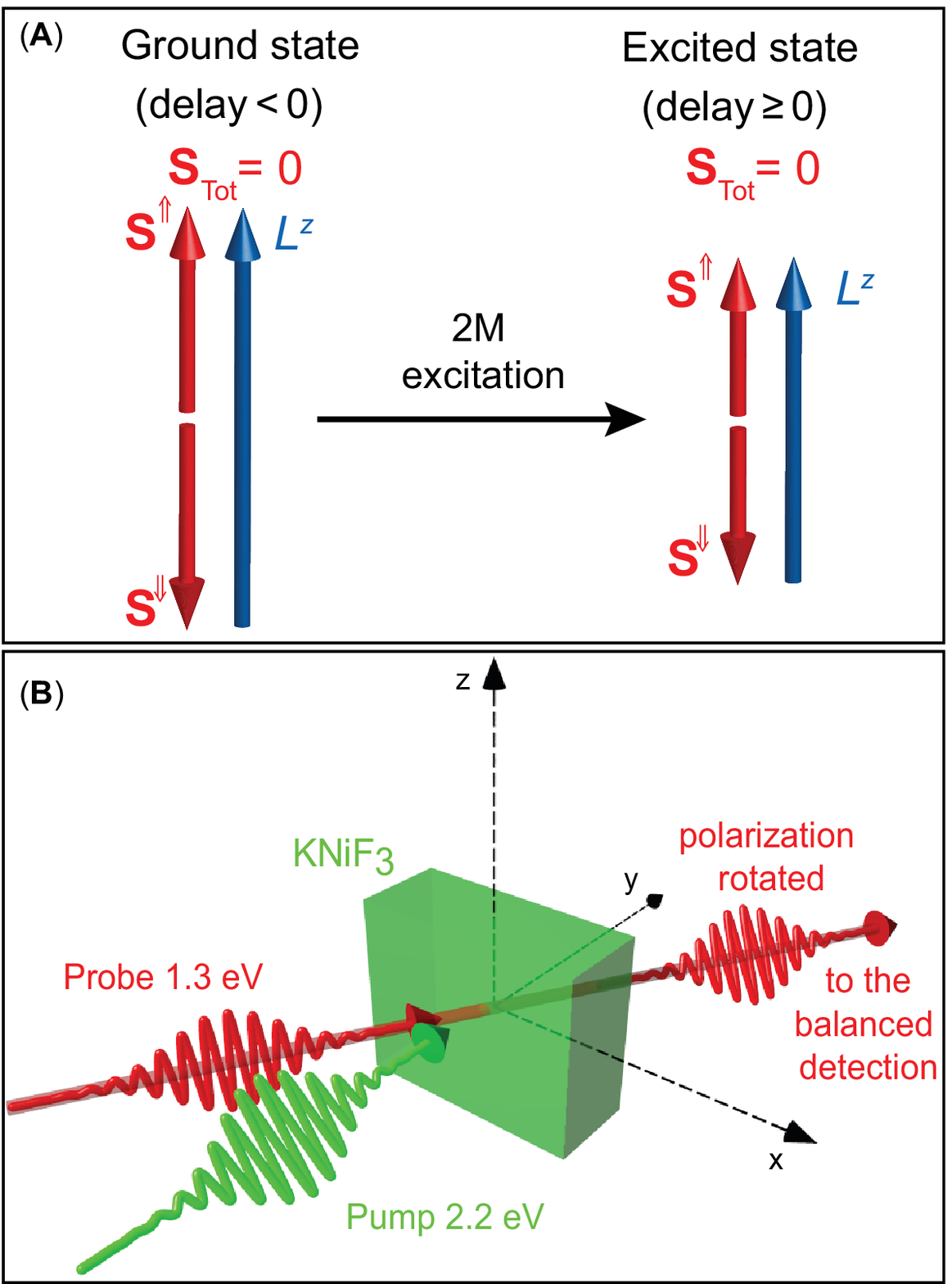}
	\caption{\footnotesize{ {\bfseries{Two-magnon mode and experimental configuration.}} (\textbf{A}) The two-magnon excitation is equivalent to a spin flip event per sublattice. Thus the magnetisation of each sublattice ($\boldsymbol{S}^{\Uparrow} ,\boldsymbol{S}^{\Downarrow}$) and, therefore, the antiferromagnetic vector ($\boldsymbol{L}_z$) is decreased in the excited state. The sum of the spins of the two sublattices, thus the total magnetisation, vanishes both in the ground and in the excited state. (\textbf{B}) Schematic representation of the experimental geometry. The setup is described in details in the Methods section.}}
	\label{fig:Fig1}
	\end{figure}

The macroscopic magnetic order of an ideal Heisenberg antiferromagnet is conveniently described in terms of the antiferromagnetic vector $\boldsymbol{L}$, which is the order parameter\cite{Landau1981} and is defined as 

	\begin{equation}
	\boldsymbol{L}= \sum_{i} \langle \hat{\boldsymbol{S}}^{\Uparrow}_{i}\rangle - \sum_{j}\langle\hat{\boldsymbol{S}}^{\Downarrow}_{j}\rangle = \boldsymbol{S}^{\Uparrow} - \boldsymbol{S}^{\Downarrow} ,
	\end{equation}
	
\noindent where $\hat{\boldsymbol{S}}^{\Uparrow}_{i}$ and $\hat{\boldsymbol{S}}^{\Downarrow}_{j}$ are the spin operators located on two nearest-neighbor sites ($i,j$), belonging to different magnetic sublattices ($\Uparrow$ and $\Downarrow$), while $\boldsymbol{S}^{\Uparrow}$ and $\boldsymbol{S}^{\Downarrow}$ are the total spins of the two sublattices [see Fig. \ref{fig:Fig1}(\textbf{A})]. In the approximation of non-interacting magnons we determined analytically that the dynamical response of the $z$-projection of $\boldsymbol{L}$ to the impulsive excitation of the 2M mode has the form

	\begin{equation}
	\Delta L^z(t) \propto A \sin{( \omega_{2M} t)} + B \cos{( \omega_{2M} t)}
	\label{eq:L}
	\end{equation}
	
\noindent where $z$ is the direction parallel to the spins, $\omega_{2M}$ is the frequency of the 2M mode and $A$ and $B$ are amplitudes. Equation (\ref{eq:L}) describes a purely longitudinal, non-precessional, dynamics of the antiferromagnetic vector (see Supplementary Materials for the complete derivation). Note that the light scattering by a magnon pair can be visualized as two spin flip events, one on each sublattice, such that the total spin remains unchanged \cite{Cottam1986,Fleury1968,Chinn1971} [see Fig. \ref{fig:Fig1}(\textbf{A})]. Consequently the transient magneto-optical Faraday and Kerr effects, which measure light-induced variations of the total spin, inevitably fail to track the dynamics of such magnetic excitation. On the other hand, the 2M process is expected to be revealed by second order magneto-optical effects, which depend on quadratic combinations of the spin operators via the same spin correlation function\cite{Ferre1984} appearing in the Heisenberg term of the Hamiltonian (see Methods). Our model reveals that the spin correlation function has the same time dependence of the antiferromagnetic vector (see Supplementary Eq. S47). The time evolution of these two quantities is unraveled by the transient antiferromagnetic linear dichroism (see Eq.(\ref{eq:epsilon})). This magneto-optical effect induces a rotation of the probe polarization in the experimental configuration shown in Fig. \ref{fig:Fig1}(\textbf{B}) (see Methods).

An excellent system for the all-optical excitation and detection of the dynamics of high-frequency and shortest-wavelength magnons is the cubic Heisenberg antiferromagnet KNiF$_3$, which is ordered below the N\'{e}el point T$_N$ = 246 K. A recent study of the dynamics of the low-energy magnons revealed that it is indeed possible to access the spin dynamics in KNiF$_3$ via a transient quadratic magneto-optical effect\cite{Bossini2014}. Moreover, in this material the Raman cross section of the 2M mode is so high that it dominates the whole spectrum\cite{Chinn1971,Cottam1986}. For the ultrafast excitation of the 2M mode in KNiF$_3$ ($\nu_{2M}\approx 22$ THz, period $\approx 45$ fs, wave vector $\approx 10^{7}$ cm$^{-1}$, wavelength $\approx 1$ nm) we rely on the ISRS mechanism to trigger a Raman-active collective mode, provided that the duration of the stimulus is shorter than the period of the mode\cite{Kimel2005a,Yan1985}. A successful impulsive excitation of such high-frequency magnons therefore demands laser pulses with a duration significantly shorter than 40 fs. To meet these requirements we used linearly polarized sub-20 fs laser pulses, with a central photon energy of 2.2 eV, which lies in the transparency window of the material \cite{Bossini2014}. For the probe we employed equally short pulses centred around 1.3 eV and with a polarization perpendicular to that of the pump.

\begin{figure}[t]
	\hspace{-1.5cm}
	\includegraphics{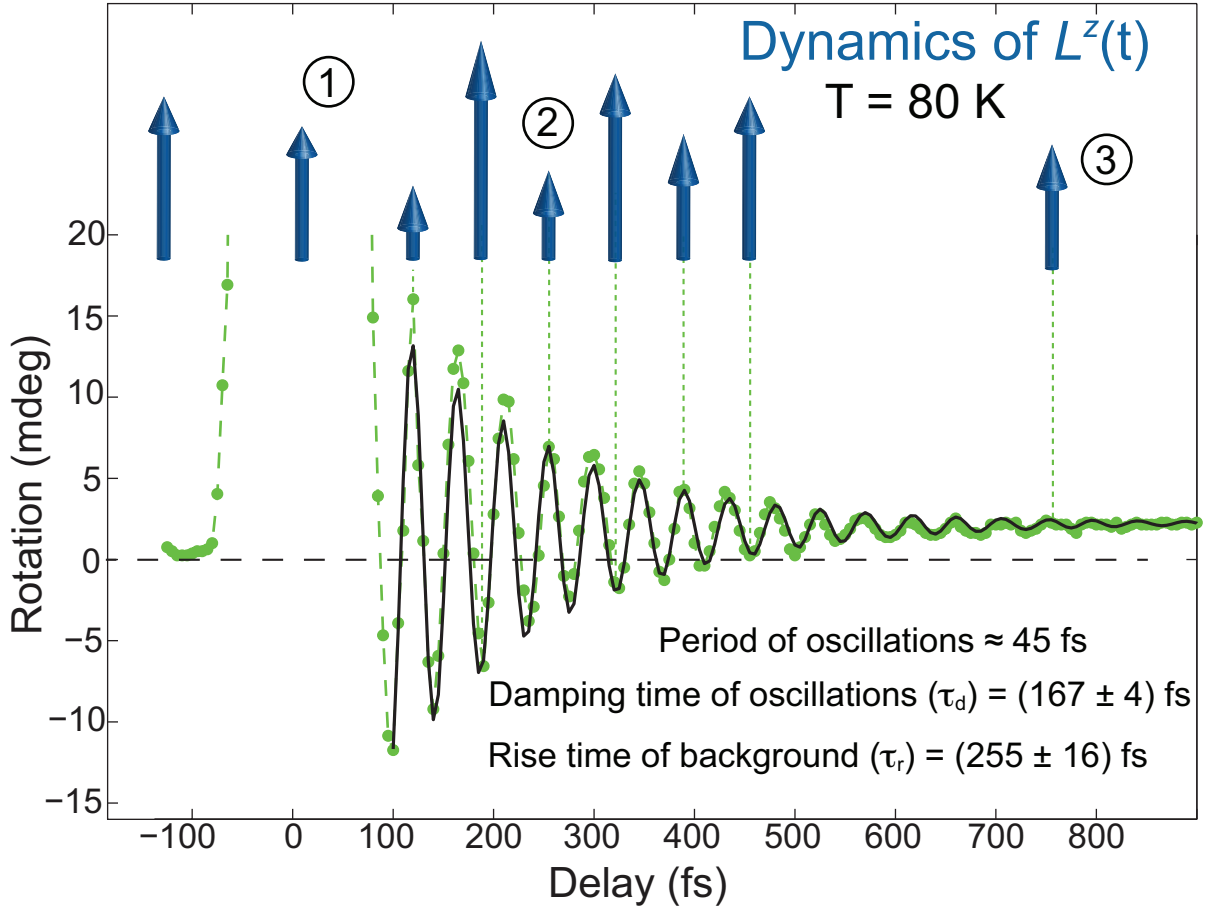}
	\caption{\footnotesize{ {\bfseries{Laser-induced dynamics of the antiferromagnetic vector.}} The transient rotation of the probe polarization was measured with the electric fields of the pump and the probe beams linearly polarized  along the $z$ and $x$ axes, respectively. The pump fluence was set to $\approx 8.6$ mJ/cm$^2$. The corresponding dynamics of the length of $L^z$ (blue arrows) is schematically represented. When the pump pulses impinge on the sample $L^z$ decreases (1), as shown in Fig. \ref{fig:Fig1}(\textbf{A}). At positive delays, oscillations at the frequency of the 2M mode are visible (2) [see Eq. (\ref{eq:L})]. The black line is a fit to the data (see the main text and the Methods section). The background contribution to the signal (3) could be interpreted as the demagnetisation of the two sublattices, responsible for the reduction of $L^z$.}}
	\label{fig:Fig2}
	\end{figure}

Figure \ref{fig:Fig2} shows the typical result of a time-resolved measurement of the laser-induced spin dynamics. The transient rotation of the probe polarization shows oscillations in time with a period of $\approx$ 45 fs (i.e. a frequency of $\approx$ 22 THz) that are damped on a 500 fs timescale. The oscillatory dynamics is superimposed on an incoherent increase of the background, as it is clear from the difference between the time trace and the zero line at longer delays (> 500 fs). To assess the nature of the 22 THz mode, we compared the temperature dependence of the time-domain signal with that of the spontaneous Raman spectra of the 2M bound state. Supplementary Figure S2 shows that the frequency and the lifetime of the pump-induced oscillations decrease as the N\'{e}el point is approached, in qualitatively and quantitatively agreement with spontaneous Raman data \cite{Chinn1971,Cottam1986}. Thus Fig. \ref{fig:Fig2} reveals the femtosecond spin dynamics triggered by the impulsive excitation of the 2M mode in KNiF$_3$, which is not accessible with any other experimental approach. A fit to the data in Fig. \ref{fig:Fig2} (see Methods) gave $\tau_d= (167 \pm 4)$ fs (damping of the coherent oscillations) and $\tau_r= (255\pm16)$ fs (rise-time of the incoherent background response). While $\tau_d$ represents the decoherence of the 2M band, we interpret $\tau_r$ as the characteristic demagnetisation time of the two sublattices, solely driven by magnetic interactions\cite{Bossini2014}. As a laser pulse excites a continuum of magnons with different frequencies [see Fig. \ref{fig:Fig5}(\textbf{A})], the damping $\tau_d$ of the oscillations observed in our experiment is actually the decoherence of the inhomogeneous ensemble of the coherently excited magnons. This is usually described \cite{Dyakonov2008} by means of the characteristic time $T_2^{\ast}$. The demagnetization of the sublattices is a result of the heating of spins, which is caused by the decoherence of single magnon modes\cite{Bossini2014} in the ensemble, on a time scale generally indicated with $T_2$ ($T_2 >T_2^{\ast}$)\cite{Dyakonov2008}. 

\begin{figure}[h]
	\includegraphics{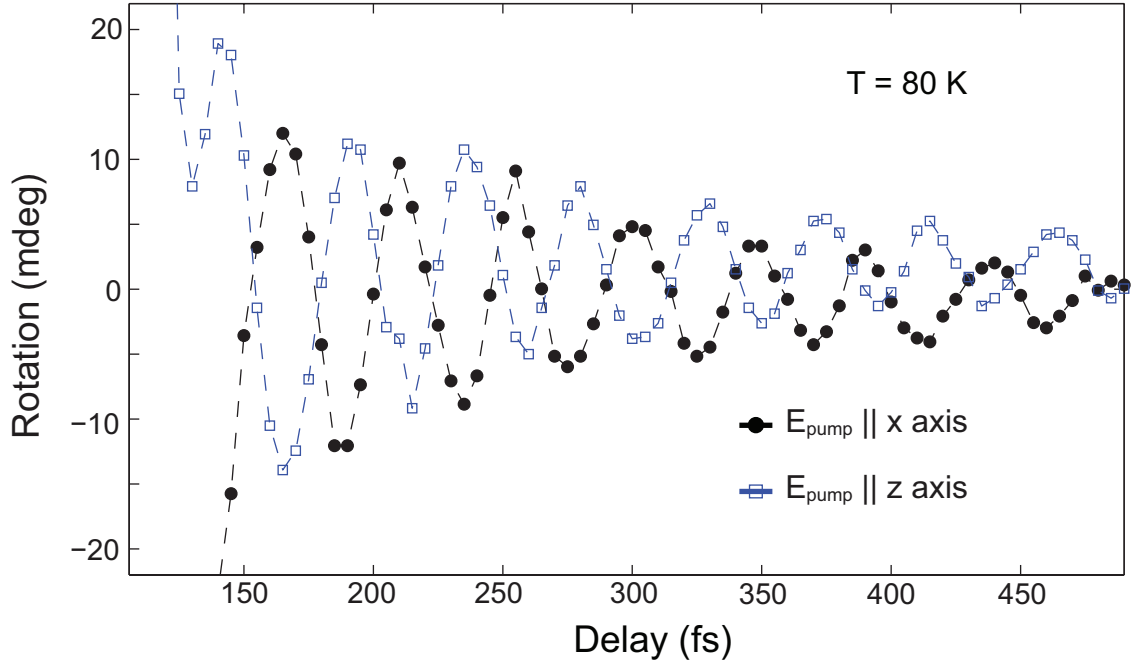}
	\caption{\footnotesize{ {\bfseries{Polarization dependence of the coherent spin dynamics.}} The phase of the oscillations is reversed by $\pi$ when the direction of the electric field of the pump beam is rotated by 90$^{\circ}$. In both the measurements the probe beam was linearly polarized along the z axis. The fluence is set to $\approx$ 12 mJ/cm$^2$ when the pump polarization is parallel to the $z$ axis, while it is $\approx$ 9 mJ/cm$^2$ in the other measurement.}}
	\label{fig:Fig4}
	\end{figure}

\begin{figure}[t]
	\includegraphics{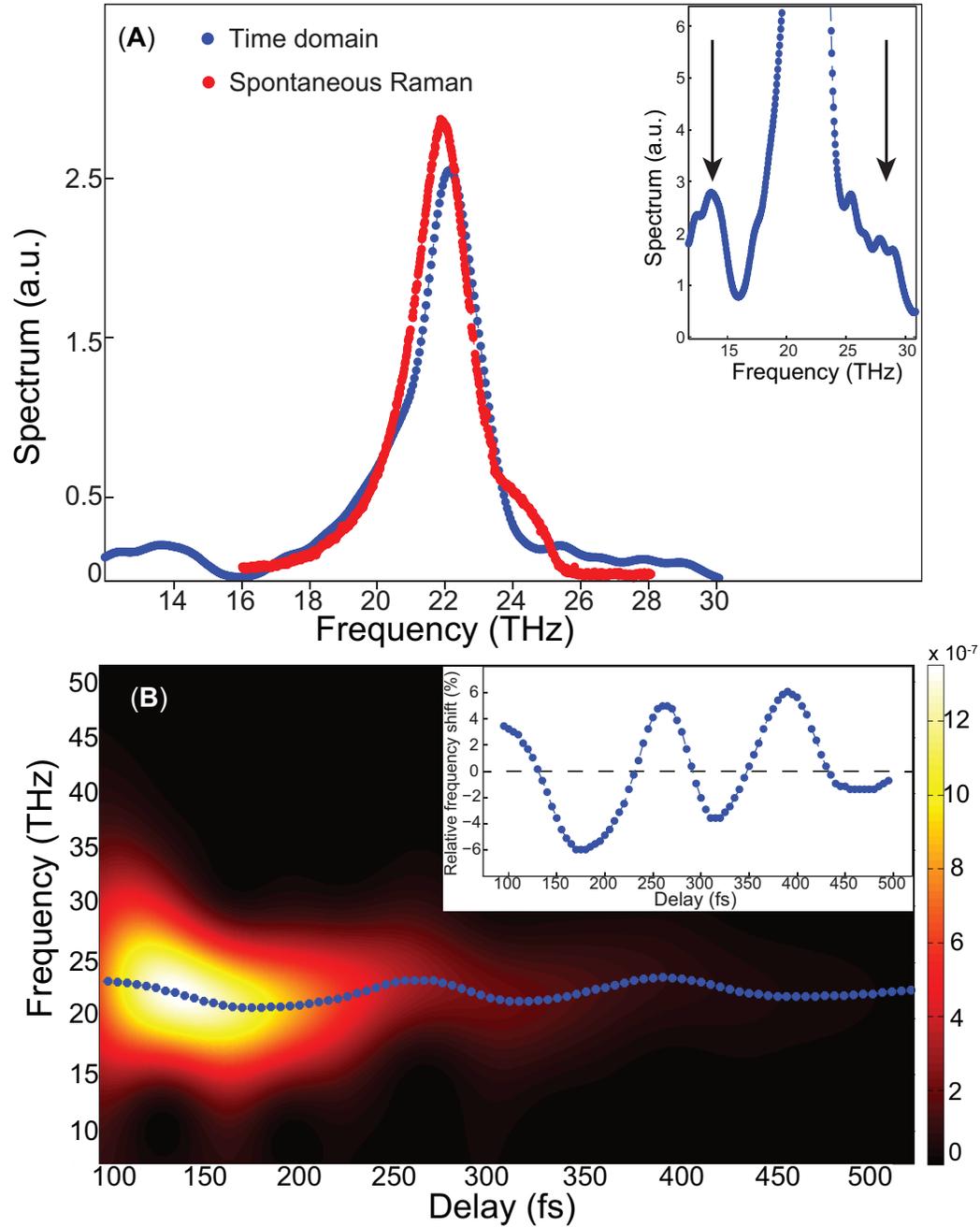}
	\caption{\footnotesize{ {\bfseries{Spectrum of the ultrafast response of the antiferromagnetic vector.}} (\textbf{A}) The Fourier transform of the time trace in Fig. \ref{fig:Fig2} measured at 80 K is compared to the spontaneous Raman spectrum obtained at the same temperature. The small discrepancy between the two lineshapes is discussed in the Supplementary Materials. In the inset a zoom of the Fourier transform reveals two sidebands at $\approx \pm7.5$ THz away from the peak frequency. (\textbf{B}) The squared modulus of the Wigner distribution of the signal is represented by the color plot. At each time step we highlight the maximum of the spectrum (blue-dotted line). In the inset the relative frequency shift [Eq. (\ref{eq:RelFreq})] is plotted as a function of the delay showing $\approx$ 7.5 THz oscillations.}}
	\label{fig:Fig5}
	\end{figure}

Moreover, only our time-resolved technique allows to observe and control the phase of the coherent short-range spin excitation, which is claimed to be a necessary requirement for any implications in the development of magnon-based devices\cite{Chumak2015}. In Fig. \ref{fig:Fig4} we plot measurements performed with orthogonal polarizations of the pump beam.  A clear $\pi$ shift of the phase of the oscillations is observed, if the polarization of the excitation beam is rotated by 90$^{\circ}$. Consequently the impulsive laser excitation of the 2M mode in KNiF$_3$ provides a phase controllable signal. Note that only a time-domain approach can reveal this feature of the light-spin interaction. This observation proves the feasibility of the coherent control of magnons in antiferromagnets near the edges of the Brillouin zone, similar to what has been realised at the zone centre\cite{Kanda2011}.

The real time measurement of the ultrafast spin dynamics triggered by short-range magnons allows us to disclose another phenomenon not observable with other techniques. Figure \ref{fig:Fig5}(\textbf{A}) shows the spectrum of the time trace in Fig. \ref{fig:Fig2} obtained by a Fourier transform (blue curve). On the same graph we plot the spectrum of the 2M mode measured via spontaneous Raman scattering at the same temperature (red curve). The zoom in the inset of Fig. \ref{fig:Fig5}(\textbf{A}) shows two sidebands at about 7.5 THz from the central 2M frequency, that appear to originate from a modulation of the 2M mode in the time domain. Figure \ref{fig:Fig5}(\textbf{B}) shows a two-dimensional spectrogram, obtained by performing a time-frequency analysis\cite{Bartelt1980,Gambetta2006} of the data in Fig. \ref{fig:Fig2} (see Supplementary Materials). The colormap in Fig. \ref{fig:Fig5}(\textbf{B}) represents the time-dependent spectrum of the 2M mode. The frequencies of the peak of the spectrum at different time delays are traced by a blue dotted line. This curve displays a periodic oscillation of $\nu_{2M}$, which is consistent with the observation of the sidebands in the inset of Fig. \ref{fig:Fig5}(\textbf{A}). We define the relative frequency shift as

	\begin{equation}
	\frac{\Delta \nu_{2M}}{\nu_{2M}}=\frac{\nu_{2M} - \langle{\nu}_{2M}\rangle}{\langle\nu_{2M}\rangle}
	\label{eq:RelFreq}
	\end{equation}\
	
\noindent where $\langle {\nu}_{2M} \rangle$ is the average frequency in the temporal interval where the oscillations have a significant amplitude (0-500 fs). We plot the peaks of the spectra obtained by the time-frequency analysis as a function of the delay in the inset of Fig. \ref{fig:Fig5}(\textbf{B}). The frequency of the modulation of $\nu_{2M}$ is $\approx 7.5$ THz, which corresponds to the frequency of the infrared-active phonon \cite{Sintani1968,Tomono1990} ($\approx 7.7$ THz), assigned to the stretching vibration of the Ni-F-Ni bond. Unlike other phonon modes in this material, the frequency of the stretching mode is temperature-independent \cite{Sintani1968,Tomono1990}, which is consistent with the data (see Fig. S6). Though this stretching mode is not Raman active in the lowest order of the electric field of light\cite{Long2002}, considering light-matter interaction at the next order (hyper-Raman scattering\cite{Long2002,Cyvin1965}) allows to excite lattice vibrations with the symmetry of the stretching mode ($F_{1u}$) in cubic crystals \cite{Cyvin1965}. We assign the modulation of $\nu_{2M}$ to the interaction on the femtosecond time-scale between the stretching mode and the 2M mode\cite{Gambetta2006} , which are simultaneously and coherently excited by the laser pulse. Although a 2M-phonon interaction was previously suggested\cite{Cottam1972}, our time-resolved experiment provides the first evidence of this effect.

Unlike previous investigations of the ultrafast spin dynamics in KNiF$_3$\cite{Bossini2014,Batignani2015}, the data reported in Fig. \ref{fig:Fig2} are a measurement of the femtosecond dynamics of the spin correlation function(see Eq.(\ref{eq:epsilon})), which carries informations about magnetic interactions on the sub-nanometer length-scale. In fact the magnetism of KNiF$_3$ is properly described by taking into account only the exchange interaction between nearest-neighbours. These antiferromagnetically coupled spins are separated approximately by an 8 {\AA} distance \cite{Lines1967}. Our experimental approach constitutes a unique way to access the ultrafast dynamics of the spin-spin correlations on such a sub-nanometer length scale. Hence we believe that this work opens up fundamentally novel and exciting perspectives for studies of magnetic and correlated materials. Following our approach it becomes possible to monitor the evolution of the exchange energy during a photo-induced phase transition and to probe the femtosecond dynamics of the sub-nanometer range spin correlations in strongly correlated materials, included high-Tc superconductors.

Although our investigation concerned an ideal Heisenberg antiferromagnet, the concept here employed to study the femtosecond dynamics of the macroscopic magnetic order parameter caused by short-range spin excitations is applicable to a broad group of multisublattice systems\cite{Fleury1968,Cottam1986}. In our view these results provide a fundamentally new approach to elucidate the dynamical interplay between short-range spin excitations and high-Tc superconductivity in cuprates \cite{LeTacon2011,DalConte2012,DalConte2015a}. 

\vspace{0.3cm}
{\large{{\bfseries{Acknowledgements}}}}

{\footnotesize{The authors thank Dr. A. Caretta and Prof. P.H.M. van Loosdredcht for the spontaneous Raman measurements. The authors thank Prof. R. Merlin, Dr. C. Giannetti and Dr. J. Mentink for fruitful discussions. This research was supported by LASERLAB-EUROPE (grant agreement n° 284464, EC's Seventh Framework Programme), de Nederlandse Organisatie voor Wetenschappelijk Onderzoek (NWO), de Stichting voor Fundamenteel Onderzoek der Materie (FOM), the European Union's Seventh Framework Program (FP7/2007-2013) Grants No. NMP3-LA-2010-246102 (IFOX), No. 280555 (Go-Fast), No. 281043 (FEMTOSPIN),  CNECT-ICT-604391 (Graphene Flagship), the European Research Council under the European Union's Seventh Framework Program (FP7/2007-2013)/ERC Grant Agreement No. 257280 (Femtomagnetism), as well as the program "Leading Scientist" of the Russian Ministry of Education of Science (14.Z50.31.0034). R.V.P. acknowledges the partial support by Russian grants 14.B25.31.0025 and 15-02-04222.}}

\vspace{0.3cm}
{\large{{\bfseries{Methods}}}}

{\footnotesize{{\bfseries{Sample.}} Our sample was a 340 $\mu$m thick (100) plane-parallel plate of KNiF$_3$, which has a perovskite crystal structure (point group $m3m$). Two equivalent Ni$^{2+}$ sublattices are antiferromagnetically coupled below the N\'{e}el temperature $T_\mathrm{N} = 246$ K\cite{Bossini2014}. This material is known to be a cubic Heisenberg antiferromagnet because of its very weak anisotropy. The positive sign of the cubic magnetic anisotropy constant determines the alignment of spins along the [001], [010] or [100] axes\cite{Landau1981}. The measurements on the KNiF$_3$ sample are carried out at a minimum temperature of 77 K in a liquid nitrogen cryostat. The temperature of the sample is monitored by a thermocouple placed on the sample holder.

{\bfseries{Light source.}} For the pump-probe experiments we used a regeneratively amplified mode-locked Ti:Sapphire laser, providing 150-fs, 500-$\mu$J pulses at 780 nm and 1 kHz repetition rate. The laser drives two Non-collinear Optical Parametric Amplifiers (NOPAs) operating in two different spectral ranges\cite{Brida2010}. Both NOPAs are pumped by the second harmonic of the laser (i.e. 390 nm) and seeded by the white-light continuum produced by focusing the 780 nm beam into a sapphire plate. The amplified pulse from the first NOPA, which initiates the dynamics (pump), has a spectrum spanning the 500-700 nm range and is compressed to nearly transform-limited duration (i.e. 8 fs) by a pair of custom-made chirped mirrors. The amplified pulse, generated by the second NOPA (probe), covers the frequency range between 820 nm and 1050 nm and is compressed to nearly  transform-limited duration (i.e. 13 fs) by a couple of fused silica prisms. 
The temporal resolution of the setup has been characterized by the cross-correlation frequency-resolved optical gating (XFROG) technique and was below 20 fs\cite{DalConte2015a}. The pump and probe beams were focused on the sample by a spherical mirror down to approximately 100 $\mu$m and 70 $\mu$m spot sizes, respectively. The high temporal resolution is preserved by using a very thin (200 $\mu$m) fused silica window as optical access to the cryostat. 

{\bfseries{Detection.}} The rotation of the polarization of the probe beam monitors the transient  antiferromagnetic linear dichroism of the sample. This magneto-optical effect is defined by the symmetric part of the dielectric tensor $\epsilon^{\lambda \nu}_{s}$ for which $\epsilon^{\lambda \nu}_{s}=\epsilon^{\nu \lambda}_{s}$, where $\nu \mbox{ and } \lambda$ are indices. The following definition holds\cite{Landau1981,Ferre1984}

	\begin{align}
	\epsilon^{\lambda \nu}_{s} = \sum_{ij} \sum_{\gamma \delta} \rho^{\lambda \nu \gamma \delta} \langle \hat{S}_{i}^{\gamma \Uparrow} \hat{S}_{j}^{\delta \Downarrow} \rangle
	\label{eq:epsilon}
	\end{align}

\noindent where $\rho^{\lambda \nu \gamma \delta}$ is a magneto-optical polar fourth rank tensor and $\lambda \nu \gamma \delta$ indicate the indices. It can be shown that the spin correlation function in the $z$ direction $\sum_{ij} \langle \hat{S}^z_i \hat{S}^z_j \rangle$ has the same time dependence as $L^z$ in Eq. (2) of the main text , which describes a solely longitudinal, non-precessional, dynamics of the antiferromagnetic vector [see Eq.(S47) of the Supplementary Materials]. Therefore the dynamics of $L^z$ must affect $\epsilon^{\lambda \nu}_{s}$ [see Eq.(\ref{eq:epsilon})] and it can be detected via the transient antiferromagnetic linear dichroism, which consists of a different absorption for light beams linearly polarised along and orthogonally to the direction of the antiferromagnetic vector. This results in the detected rotation of the probe polarization. We measured the pump-induced rotation of the probe polarization employing a balanced-detection scheme. The transmitted probe is split by a Wollaston prism into two orthogonal linearly polarized beams and focused on a couple of balanced photodiodes. The Wollaston prism is rotated in order to equalize the probe intensities on the two photodiodes. The pump-induced imbalance of the signal registered by the two photodiodes is measured by a lock-in amplifier which is locked to the modulation frequency of the pump beam (i.e. 500 Hz). Our apparatus was able to detect rotations of the polarization on the order of 1 mdeg.

{\bfseries{Fitting procedure.}} The transient rotation of the probe polarization was fitted employing the following function

	\begin{equation}
	\nonumber
	f(t) = C \sin{(\tilde{\omega}t + \phi)}e^{(-t/\tau_d)} + H(t)D(1-e^{(-t/\tau_r)})
	\end{equation}

\noindent where $C$ and $D$ are amplitude coefficients, $\phi$ is the phase of the oscillations, $H(t)$ is the Heaviside function, $\tau_d$ is the damping time of the oscillations and $\tau_r$ is the characteristic rise time of the incoherent contribution to the signal. Considering the outcome of the Wigner analysis in Fig. 4(\textbf{B}), we employed the modulated frequency  $\tilde{\omega}$ in the sinusoidal function, namely

	\begin{equation}
	\nonumber
	\tilde{\omega} = 2 \pi \nu_{2M} (1 + G \sin{( 2 \pi \nu_{mod}t + \psi)})
	\end{equation}

\noindent where $\nu_{2M}$ is the 2M frequency, $G$ is an amplitude coefficient, $\nu_{mod}$ is the modulation frequency and $\psi$ is the phase. The fit to the data shown in Fig. 2 was achieved by setting the following parameters: $\nu_{2M} = 22.12$ THz, $G=0.002$, $\nu_{mod} = 7.5$ THz, $\psi=45^{\circ}$, $D=2.3\cdot10^{-3}$ deg. These values were obtained from the Wigner analysis of the data in Fig. 2. The fit parameters allowing to reproduction at best the data are: $C=(2.5 \pm 0.1)\cdot10^{-2}$ deg, $\phi=(220\pm1)^{\circ}$, $\tau_d = (167 \pm 4)$ fs, $\tau_r= (255\pm16)$ fs.}}

%


\end{document}